\documentclass[a4paper,11pt]{article}
\pdfoutput=1 

\usepackage{jheppub} 

\usepackage[T1]{fontenc} 
\usepackage[english]{babel}
\usepackage{ulem}
\usepackage{xcolor}
\usepackage{amsmath}
\usepackage{tensor}
\usepackage[utf8]{inputenc}
\usepackage{cleveref}
\usepackage{graphicx}
\usepackage{float}
\usepackage{subfig}

\definecolor{alyssapink}{RGB}{250, 50, 150}

\title{Shift-Type SMEFT Effects in Dileptons at the LHC}
\author{Alyssa Horne,}
\author{Jordan Pittman,}
\author{Marcus Snedeker,}
\author{William Shepherd,}
\author{and Joel W. Walker}

\emailAdd{adh070@shsu.edu}
\emailAdd{jip003@shsu.edu}
\emailAdd{mks039@shsu.edu}
\emailAdd{shepherd@shsu.edu}
\emailAdd{jwalker@shsu.edu}

\affiliation{Physics Department, Sam Houston State University, Huntsville TX 77341, USA}

\abstract{We explore the constraints which can be derived on Wilson coefficients in the Standard Model Effective Field Theory from dilepton production, notably including the constraints on operators which do not lead to cross sections growing with energy relative to the Standard Model rate, i.e. shifts. We incorporate essential theory error estimates from higher EFT orders in the analysis in order to provide robust bounds. We find that constraints on four-fermion operator contributions which do grow with energy are not materially weakened by the inclusion of these shifts, and that a constraint on the shifts can also be derived, with a characteristic strength comparable to, and a directionality in parameter space complementary to, those from LEP data. This completes the study of hadronically-quiet dilepton production in the SMEFT, and provides two new constraints which are linearly independent from others arising at the LHC and also rotated in Wilson coefficient space relative to, though not completely independent from, the LEP bounds.}

\begin{document}

\maketitle

\section{Introduction}

The LHC has generated extensive amounts of data, and has made the Standard Model (SM) paradigm-affirming discovery of the Higgs boson~\cite{Chatrchyan:2012ufa,Aad:2012tfa}. Building on the already-impressive performance of the machine, detectors, and experimental collaborations, the schedule calls for an increase by more than an order of magnitude in the integrated luminosity available for detailed investigation over the coming 10-15 years. This unprecedented quantity of data at the highest energies we can probe will enable incredibly precise measurements that will have unique sensitivity to new dynamics.

Sadly, the copious production of new particles, so broadly anticipated in the ramp-up to the LHC, has failed to materialize. Current $2\sigma$ constraints on popular realizations of new physics (NP) are already near or beyond the expected ultimate $5\sigma$ discovery reach of the full $3\,{\rm ab}^{-1}$ dataset envisioned for the future of the LHC. Given this fact, it seems unlikely that we will be able to directly produce and detect new fundamental particles at the LHC.

A natural and intuitive response to this dual state of affairs, where we will be privileged with incredibly precise measurements at the highest energy ever attained in a laboratory experiment and yet unable to detect new particles directly in that data set, is to move toward more model-independent means of interpreting the measurements we are nonetheless able to make. The canonical tool, used throughout the history of particle physics, for understanding physics at scales well-separated from each other is Effective Field Theory (EFT). The application of EFT techniques to the situation where we seem to understand the basic structure of electroweak symmetry breaking based on the Higgs data has come to be called the Standard Model Effective Field Theory, or SMEFT \cite{Brivio:2017vri}. As long as whatever new physics we anticipate is heavy compared to the energy scales being probed and the Higgs we've discovered plays the full role of Electroweak symmetry breaking, the SMEFT provides a complete description of realizable outcomes and a uniform language for comparing data to arbitrary model hypotheses, including models which may yet be explicitly formulated far in the future.

The crucial feature of the SMEFT that enables this model-independent reorganization of effects is the perturbation series introduced in $\frac{s}{\Lambda^2}$, the ratio of the scale characterizing the experiment $\sqrt s$ to the new physics scale $\Lambda$. The combination of otherwise distinct coupling orders in the NP interactions to instead partition by powers of mass suppression is the primary feature of the EFT approach. This leads to a finite (albeit large) number of leading-order operators, first properly enumerated and listed by~\cite{Grzadkowski:2010es}. With more recent techniques, it is possible to count the number of independent operators at any given order in $\frac{1}{\Lambda^2}$~\cite{Henning:2015alf}; results indicate that the number of independent parameters (in a model-independent treatment) grows very rapidly with the order of the expansion parameter, sharply limiting our ability to perform such a model-agnostic analysis at beyond leading-order in the EFT perturbation series; so far we lack the appropriate number of independent constraints on the SMEFT already at leading order, so introducing more parameters which need constraining will only make the problem worse.

However, our inability to consistently constrain all operators at dimension 8 (NLO in the EFT perturbation expansion) does not justify assuming the existence of those operators away. Just like any other calculation in perturbation theory, there can be no doubt that errors remain from uncalculated terms in the series, and treating those theoretical errors honestly and conservatively is essential to producing a result that accurately describes what we have learned from our analysis. If we instead make aggressive assumptions about the behavior of the theory at higher orders, for example neglecting these errors entirely, we will produce ``constraints'' that are overly aggressive, and thus more complete models will exist which violate our constraints while simultaneously being consistent with the data from which they have been extracted, making those constraints effectively useless to the HEP community.

One may attack the large dimensionality of the SMEFT coefficient space systematically by restricting the process under consideration.
Drell-Yan dilepton production is of particular interest at high energies because of the interplay with already impressive constraints available in pre-LHC data at lower energies from $Z$-pole and low-energy scattering processes \cite{Berthier:2015oma,Berthier:2015gja,Bjorn:2016zlr,Berthier:2016tkq,Hartmann:2016pil,Falkowski:2017pss}. This is a particularly clean final state for study at the LHC, and the potential impact on the SMEFT parameter space has been explored by many authors, with different motivations and viewpoints~\cite{Farina:2016rws,Greljo:2017vvb,Alioli:2017nzr,Alioli:2018ljm,Dawson:2018dxp,Fuentes-Martin:2020lea}. In a recent study~\cite{Alte:2018xgc}, one of us and collaborators explored the effect of theoretical uncertainties on the constraints that could be derived from these processes, considering only operators which gave contributions growing with energy. The constraints derived there were a bit weaker than those found by more aggressive methods, as expected.

In this article, we expand the types of operators whose effect we consider to include shift-type operators, which give signal effects that do not grow with energy like those previously studied. These operators are sometimes considered to have already been adequately constrained by $Z$-pole data to justify their exclusion from LHC analyses, but we will see that their inclusion leads to quantitative changes to the LHC's constraining power in this expanded parameter space. With their inclusion, the LHC has sensitivity in this final state to three distinct linear combinations of Wilson coefficients from the SMEFT; an additional three directions in parameter space contributing to dilepton production are unable to be constrained using LHC data alone.

In the next section, we will review the SMEFT formalism and explore the effect of these shift-type operators on the LHC dilepton rate. In \cref{sec:stats} we will describe the statistical framework of our study, including error estimates, and in \cref{sec:search} we will apply those methods to a search inspired by the ATLAS collaboration. In \cref{sec:results} we discuss the import of these results, how they revise our understanding of the LHC's role in constraining the SMEFT, and the ultimate future goals of the SMEFT program.

\section{SMEFT Formalism and Operators Relevant to Dilepton Production}
\label{sec:shifts}

In EFT, complex interactions involving particles whose mass is far above the dynamical scale of an experiment are approximated by point like interactions (a canonical example being Fermi's theory of weak decay). This estimation exploits the smallness of a ratio of scales, in which a perturbation expansion can be done. In the case of the SMEFT, a perturbation expansion in $\frac{E^2}{\Lambda^2}$ is the central feature, with $\Lambda$ being the NP scale. This operates under the assumption that the 125 GeV boson found by the LHC is the Standard Model Higgs boson responsible for electroweak symmetry breaking, as indicated by all current data. The full Standard Model symmetry and particle content is retained, but additional operators corresponding to higher dimensional point-like interactions built entirely out of Standard Model fields are introduced to parameterize unknown heavy NP. The SMEFT Lagrangian can be written in the form:
\begin{equation}\label{eq:lag}
\mathcal{L}_{SMEFT} = \mathcal{L}_{SM} + \mathcal{L}^{(5)} + \mathcal{L}^{(6)} +\mathcal{L}^{(7)} + \mathcal{L}^{(8)} + ...
\end{equation}
Here $\mathcal{L}_{SM}$ denotes the would-be SM Lagrangian, which consists of operators of dimension $d\leq4$; note that the parameters in these interactions can be affected by SMEFT effects in input measurements~\cite{Berthier:2015oma,Brivio:2017bnu}. The latter Lagrangian terms in \cref{eq:lag} then take the following form:
\begin{equation}
\mathcal{L}^{(i)} = \sum_{k = 1}^{N_i} \frac{C^{(i)}_k}{\Lambda^{i-4}} Q^{(i)}_k.
\end{equation}
In this sum, $C_k^{(i)}$ represents the appropriate Wilson coefficient and $Q_k^{(i)}$ the associated operator of dimension $i$. After significant effort, operator bases up to dimension 9 are known~\cite{Weinberg:1979sa,Wilczek:1979hc,Buchmuller:1985jz,Grzadkowski:2010es,Abbott:1980zj,Lehman:2014jma,Lehman:2015coa,Liao:2016hru,Li:2020xlh,Liao:2020jmn}; the dimension-5 contributions consist only of the operator responsible for Majorana neutrino masses~\cite{Weinberg:1979sa,Wilczek:1979hc}, and we use the ``Warsaw Basis''~\cite{Grzadkowski:2010es} of dimension-6 operators; for our analysis at $\mathcal{O}\left(\frac{1}{\Lambda^2}\right)$ no operators of higher dimensionality are necessary.

There are two ways in which the SMEFT can impact dilepton production. Direct SMEFT effects arise as a result of four-fermion operators, and lead to effects which grow with collision energy relative to the SM amplitude. Shift effects, by contrast, arise from redefinitions of would-be SM couplings due to SMEFT effects, either in the dilepton production process itself or in some other process which is utilized as an input measurement to define the Lagrangian parameters in terms of physical observables; these effects are proportional to the SM amplitude, and thus do not grow with energy in the same way that the direct contributions do.

\begin{table}[t]
\begin{tabular}{|c|c|c|}
Shift Operators&Direct Forward Operators&Direct Backward Operators\\\hline
$Q_{HWB}\equiv H^\dag \tau^I H\, W^I_{\mu\nu} B^{\mu\nu}$ &$Q_{lq}^{(1)}\equiv\left( \bar{l}_p \gamma_\mu l_p \right) \left( \bar{q}_s \gamma^\mu q_s \right)$ & $Q_{lu}\equiv \left( \bar{l}_p \gamma_\mu l_p \right) \left( \bar{u}_s \gamma^\mu u_s \right)$\\
$Q_{ll}^\prime\equiv(\bar l_p \gamma_\mu l_s)(\bar l_s \gamma^\mu l_p)$& $Q^{(3)}_{lq}\equiv\left( \bar{l}_p \gamma_\mu \tau^I l_p \right) \left( \bar{q}_s \gamma^\mu \tau^I q_s \right)$& $Q_{ld}\equiv \left( \bar{l}_p \gamma_\mu l_p \right) \left( \bar{d}_s \gamma^\mu d_s \right)$ \\
$Q_{Hd}\equiv(H^\dag i\overleftrightarrow{D}_\mu H)(\bar d_p \gamma^\mu d_p)$&$Q_{eu}\equiv \left( \bar{e}_p \gamma_\mu e_p \right) \left( \bar{u}_s \gamma^\mu u_s \right)$&$Q_{qe}\equiv \left( \bar{q}_p \gamma_\mu q_p \right) \left( \bar{e}_s \gamma^\mu e_s \right)$\\
$Q_{Hu}\equiv(H^\dag i\overleftrightarrow{D}_\mu H)(\bar u_p \gamma^\mu u_p)$& $Q_{ed}\equiv \left( \bar{e}_p \gamma_\mu e_p \right) \left( \bar{d}_s \gamma^\mu d_s \right)$&\\
$Q_{He}\equiv(H^\dag i\overleftrightarrow{D}_\mu H)(\bar e_p \gamma^\mu e_p)$& &\\
$Q_{Hl}^{(1)}\equiv(H^\dag i\overleftrightarrow{D}_\mu H)(\bar l_p \gamma^\mu l_p)$&  &\\
$Q_{Hl}^{(3)}\equiv(H^\dag i\overleftrightarrow{D}^I_\mu H)(\bar l_p \tau^I \gamma^\mu l_p)$& &\\
$Q_{Hq}^{(1)}\equiv(H^\dag i\overleftrightarrow{D}_\mu H)(\bar q_p \gamma^\mu q_p)$&&\\
$Q_{Hq}^{(3)}\equiv(H^\dag i\overleftrightarrow{D}^I_\mu H)(\bar q_p \tau^I \gamma^\mu q_p)$&&\\
$Q_{HD}\equiv\ \left(H^\dag D_\mu H\right)^* \left(H^\dag D_\mu H\right)$&&\\
\end{tabular}
\caption{\label{tab:ops}Table of operators contributing to dilepton production in the SMEFT, sorted into shift-type operators and direct operators, which are further sorted by angular behavior. Note that here we use the conventions of~\cite{Grzadkowski:2010es}; in particular, $p,s$ are flavor indices.}
\end{table}

In addition to these two different possible energy behaviors, there are also two possible angular behaviors that arise due to SMEFT effects. In the case of direct contributions the nature of these differences is most clear; there can be a preference for forward or backward (negatively charged) lepton production in the hard scattering, which results from angular momentum conservation for a given helicity structure. In the case of operators which shift would-be SM couplings, couplings to both helicities of quarks and/or leptons are generically altered, such that a shift operator must contribute to both forward- and backward-preferential lepton production.

Accounting for both energy and angular behaviors, there are four effects that we may potentially observe and disentangle from one another. These effects are labeled as forward, backward, shift-forward, and shift-backward. We define our forward and backward contribution parameters from the Wilson coefficients $C_{lq}^{(3)}$ and $C_{lu}$, respectively, as exemplars of the full linear combinations which give rise to their effects, originally calculated in~\cite{Alte:2018xgc},

\begin{eqnarray}
    c_{fwd}& =& C^{(3)}_{lq} - 0.48C_{eu} - 0.33C^{(1)}_{lq} + 0.15C_{ed},\\
    c_{bwd}& =& C_{lu} + 0.81C_{qe} - 0.33C_{ld}.
\end{eqnarray}

The shift operators, in addition to each being able to contribute to both forward and backward scattering, suffer an additional complication relative to the direct contributions. Unlike in the direct case, where the helicity structure is completely determined by the operator itself, there can be different angular behavior that arises due to interference with different SM graphs. Of course, the full physical effect is always due to the interference with the full SM amplitude, but the different behavior with energy of photon versus $Z$-boson graphs means that these differences are also physical, and can in principle be disentangled by considering different regimes in $\sqrt{s}$.

Given this behavior, there are in total four distinct contributions that shift operators can make to dilepton production; one enumeration would consider forward and backward scattering due to interference with $Z$ or photon SM graphs. While this may be the most straightforward set of contributions to calculate, it is not the most straightforward to measure. We instead define our four linear combinations as forward and backward contributions at high $\sqrt{s}\gg M_Z$ and at the $Z$ pole.

\begin{figure}[t]
    \center
    \subfloat{{\includegraphics[width=0.45\textwidth]{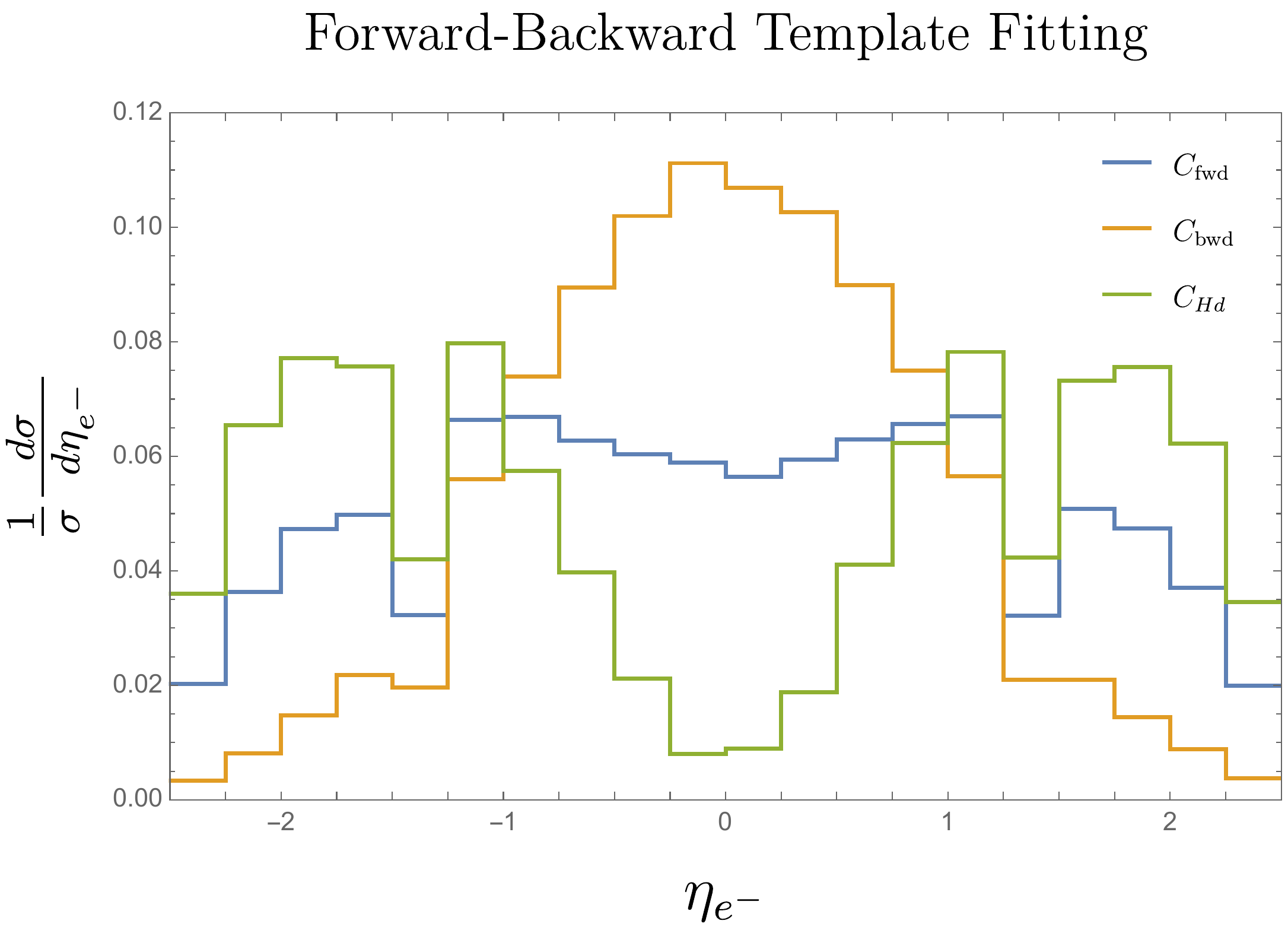}}}\hspace{1em}
    \subfloat{{\includegraphics[width=0.45\textwidth]{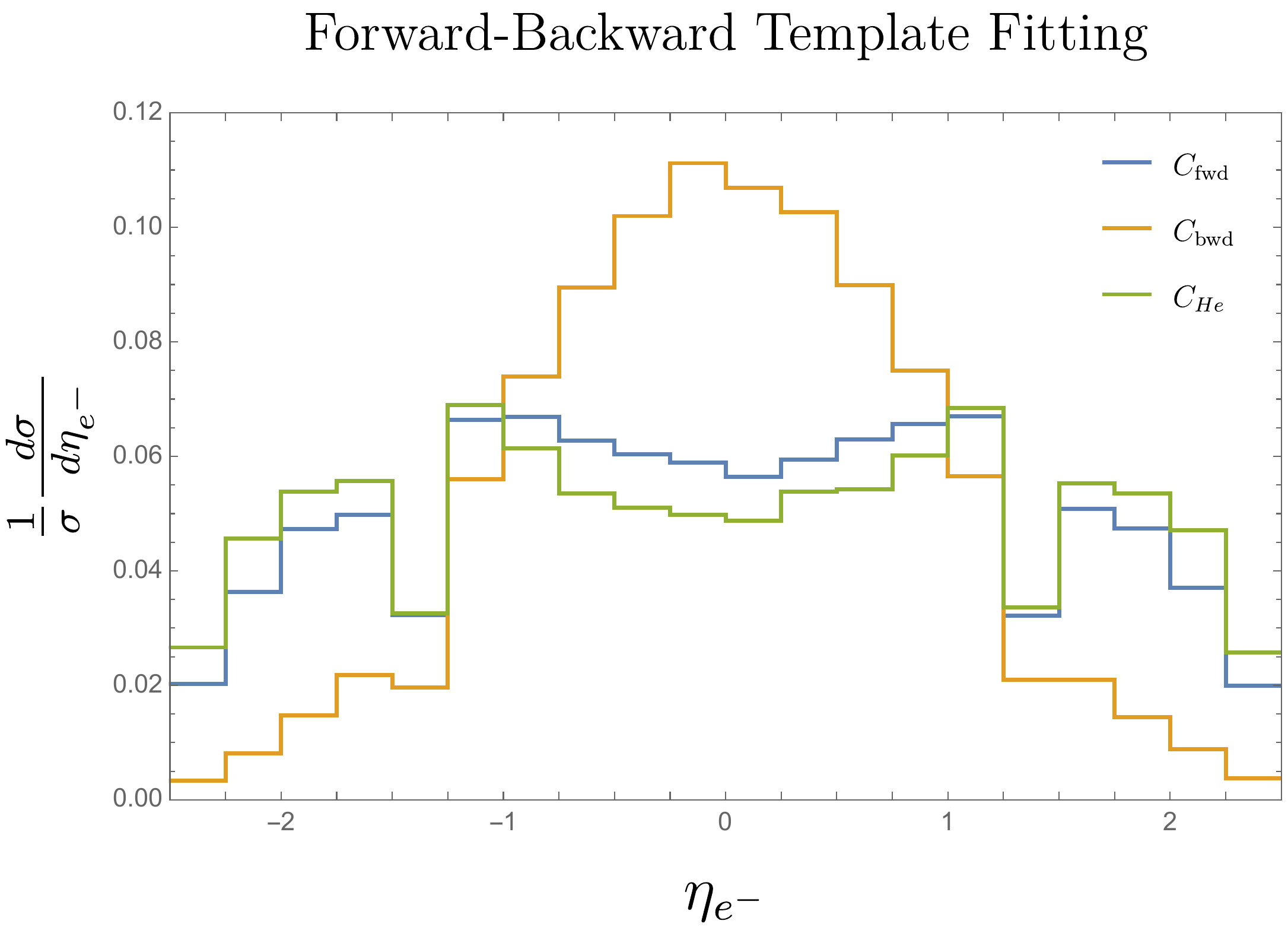}}}
    \hfill 
    
    \center
    \subfloat{{\includegraphics[width=0.45\textwidth]{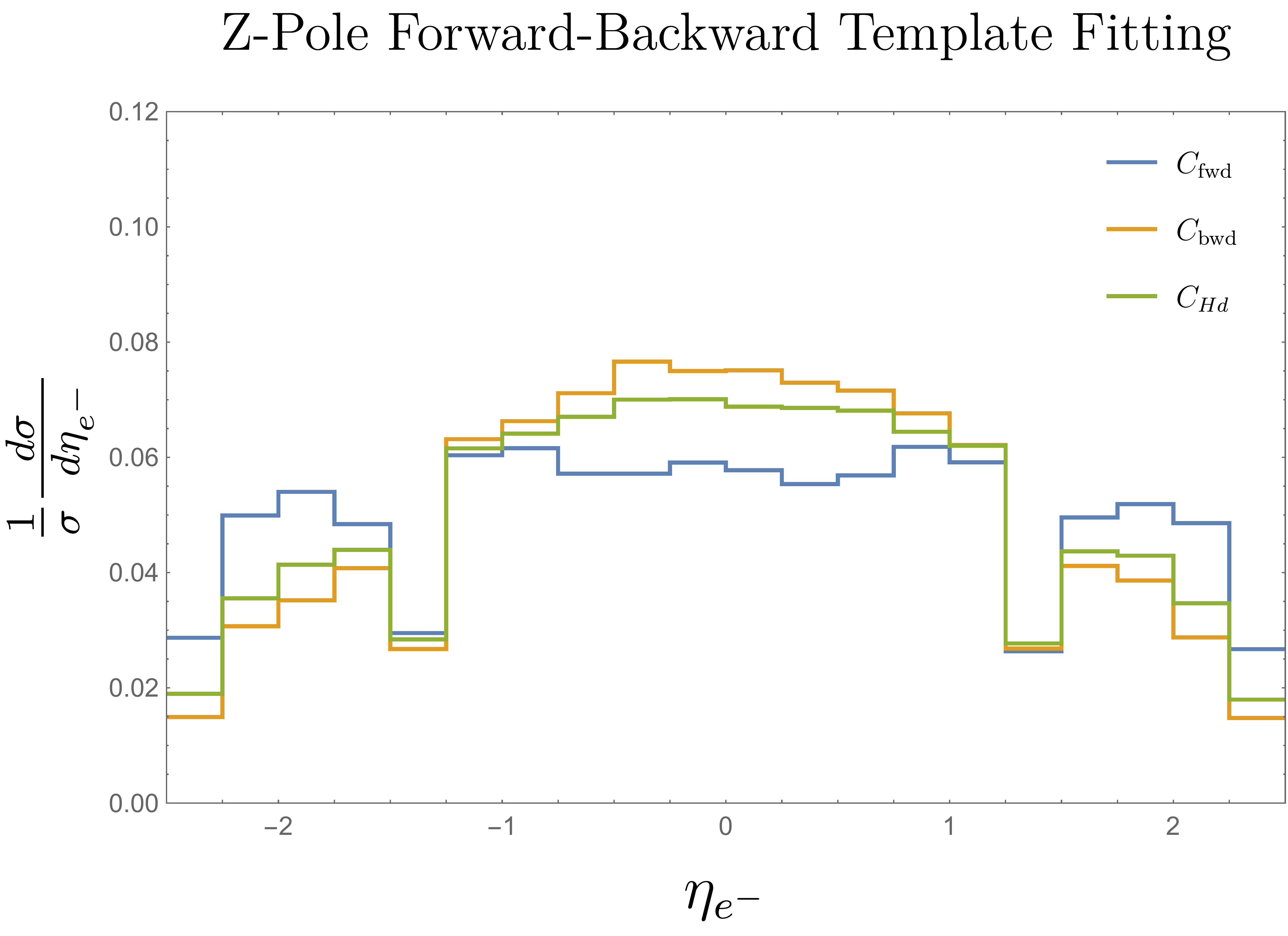}}}\hspace{1em}
    \subfloat{{\includegraphics[width=0.45\textwidth]{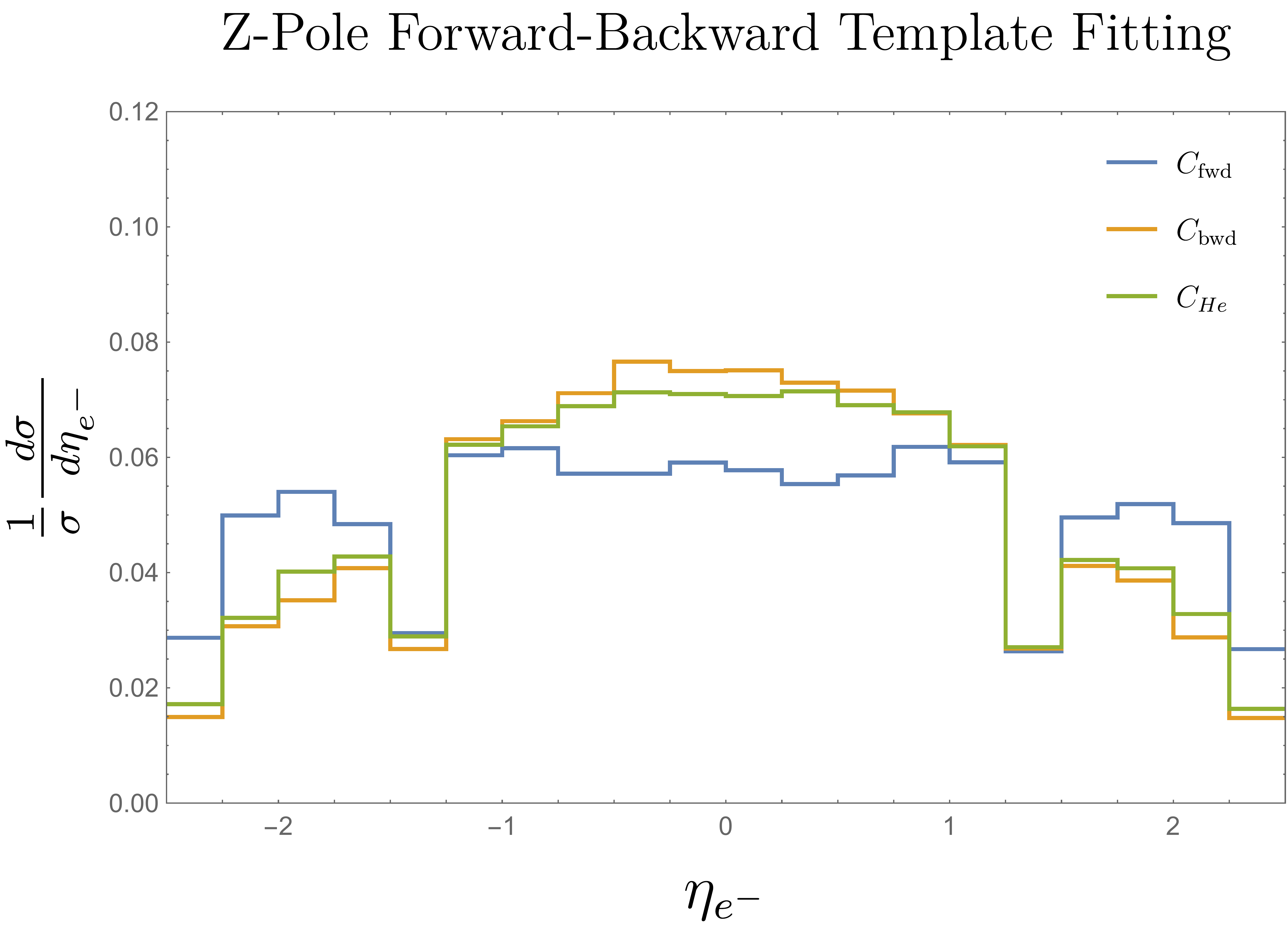}}}
    \hfill
    \caption{Examples of Template Fitting for $C_{Hd}$ (left) and $C_{He}$ (right) for high $\sqrt{s}$ (upper row) and $Z$-pole (lower row) samples.  The region between $1.37 < \vert\eta\vert < 1.52$ is excluded due to reduced energy resolution in the ATLAS detector~\cite{Aaboud:2017buh}.}
    \label{fig:templates}
\end{figure}

Simulation of the direct-contribution exemplar operators was used to generate the normalized particle pseudo-rapidity template distributions shown in \cref{fig:templates}, with blue for $c_{fwd}$ and orange for $c_{bwd}$.
For each of the shift operators in each relevant $M_{\ell\ell}$ bin, we performed a binary template fit, minimizing the bin-wise sum of square-differences relative to $\alpha\times c_{fwd} + (1-\alpha)\times c_{bwd}$.

Examples are provided for the operators $C_{Hd}$ and $C_{He}$ (green contours) in \cref{fig:templates}, respectively with $\alpha = (1.92,1.21)$ in the high $M_{\ell\ell}$ region and $\alpha=(0.32,0.20)$ in the $Z$-pole bin; the former exhibits interesting anti-correlation with the backward template, whereas the latter is predominantly forward at high $\sqrt{s}$, and the difference in behavior between the two energy regimes is very apparent. Note that the forward and backward distributions themselves become less easily distinguished at lower energies; this results from the fact that these events are less likely to probe the high momentum fraction region of the PDF, where the valence quark distributions best motivate our definition of forward and backward.

From the resulting list of coefficients $\alpha_i$ and the associated cross sections, we extracted linear combinations of Wilson coefficients which will be constrained by LHC measurements of forward and backward shift-type contributions to dilepton production, as follows:\footnote{The normalization of these linear combinations is chosen to allow our exemplar functions to have unit length in Wilson-coefficient space, assuming a trivial $\delta_{ij}$ metric.}

\begin{align}
c_{\rm fwd,hi}^{\rm(shift)}=& -6.7 C_{Hd} -54 C_{HD} -21 C_{He} +84 C_{Hl}^{(1)} -130 C_{Hl}^{(3)} -26 C_{Hq}^{(1)}\nonumber\\& +62 C_{Hq}^{(3)} +28 C_{Hu} -120 C_{HWB} +110 C_{ll}^\prime, \\
c_{\rm bwd,hi}^{\rm(shift)}=&\,\, 3.9 C_{Hd} +6.8 C_{HD} +4.6 C_{He} -16 C_{Hl}^{(1)} +10. C_{Hl}^{(3)} +23 C_{Hq}^{(1)}\nonumber\\& -7.3 C_{Hq}^{(3)} -21 C_{Hu} +5.3 C_{HWB} -13 C_{ll}^\prime.\\
c_{{\rm fwd,}Z}^{\rm(shift)}=&\,\, -4.4 C_{Hd} -49 C_{HD} -21 C_{He} +72 C_{Hl}^{(1)} -120 C_{Hl}^{(3)} -2.4 C_{Hq}^{(1)}\nonumber\\& +59 C_{Hq}^{(3)} +11 C_{Hu} -110 C_{HWB} +96 C_{ll}^\prime.\\
c_{{\rm bwd,}Z}^{\rm(shift)}=&\,\, -0.53 C_{Hd} -1.6 C_{HD} -4.9 C_{He} +2.9 C_{Hl}^{(1)} -3.6 C_{Hl}^{(3)} +0.83 C_{Hq}^{(1)}\nonumber\\& +4.4 C_{Hq}^{(3)} +0.96 C_{Hu} +1.2 C_{HWB} +3.3 C_{ll}^\prime.&
\end{align}

Given these linear combinations, we must choose exemplar cases to simulate and then constrain. Since every shift operator contributes to every behavior of interest, it is necessary to use four operators simultaneously to construct distributions which correspond to the four physically-distinct contributions of interest. We chose to construct those samples by simultaneously simulating the effects of $C_{He},\,C_{H\ell}^{(3)},\,C_{\ell\ell}^\prime$ and $C_{Hq}^{(1)}$. We will consistently label these exemplars by the linear combination which they exemplify throughout the remainder of the article.

\section{Statistical Methods and Error Estimates}
\label{sec:stats}

We use a straightforward $\chi^2$ goodness-of-fit test for statistical inference here, as we lack the detailed information necessary to use more sophisticated tools. We will report as constrained any point in parameter space where the signal model fails to be a good fit at the 95\% C.L. to the SM prediction.
Note that this is a different procedure from the $\Delta\chi^2$ test which would be used to estimate a parameter, and results in somewhat more conservative bounds.

We explored the differences in these analysis strategies by using pseudo-data generated by a Poisson distribution with the SM expectation as its mean in each bin. Testing the $\Delta \chi^2$ approach against the straightforward $\chi^2$ goodness-of-fit test, we noted that the goodness-of-fit test indeed only provides more conservative bounds, and further that even under background fluctuations the $\Delta\chi^2$ analysis never provided a bound outside that from the $\chi^2$ approach used here. Thus, regions of parameter space which we claim to constrain can reliably be expected to be constrained. Once data has been properly collected, a $\Delta\chi^2$ analysis should also constrain some portion of the ``allowed'' region according to our current analysis, but it is not possible to predict which a priori.

\subsection{Treatment of Errors in the SMEFT: General Technique}

As with any statistical examination, the essential element of the analysis is actually the treatment of errors. Statistical errors for a binned fit are always purely Poisson, with bin-by-bin variance of

\begin{equation}
    \sigma_{\rm stat}^2=N_{\rm bg},
\end{equation}
where $N_{\rm bg}$ is the SM background prediction for the number of events in the given bin.

As we are beginning from a completed ATLAS search~\cite{Aaboud:2017buh} as a template for our analysis, we are able to normalize the systematic errors of our analysis to those reported by ATLAS. Given the origin of many systematic errors from statistical measurements in control regions, we imagine that these bounds will improve with the same functional behavior as statistical errors. However, these can never be fully eliminated, so we place a floor of 2\% on the relative error on any given bin's content due to systematics\footnote{A different value for this floor would be appropriate for more intrinsically difficult-to-measure final states than dileptons}. This gives us a systematic variance contribution of

\begin{equation}
    \sigma^2_{\rm sys}=\max\left(\sigma^2_{{\rm sys, }36}\left(\frac{L_{\rm int}}{36~{\rm fb}^{-1}}\right),\left(0.02 N_{bg}\right)^2\right),
\end{equation}
where $\sigma^2_{{\rm sys, }36}$ is the ATLAS reported error on 36 fb$^{-1}$ of data.

The most novel error treatment we report follows the formalism originally developed by \cite{Alte:2017pme} and applied to dilepton processes (studying only those contributions growing with energy) in \cite{Alte:2018xgc} in explicitly estimating the SMEFT contribution at $\mathcal{O}\left(\frac{1}{\Lambda^4}\right)$ and treating that as a theoretical error in our predictions, which derives no benefit from increased event statistics.

Our estimate for the total SMEFT effect at next order in the EFT expansion is calculated by taking the square of the $\mathcal{O}\left(\frac{1}{\Lambda^2}\right)$ amplitude as a template for the generic size of effects at $\mathcal{O}\left(\frac{1}{\Lambda^4}\right)$, scaling it by a factor to account for the additional $N_8$ dimension-8 operators which will contribute\footnote{This parameter of the analysis must be estimated; we take $N_8=20$ throughout to be concrete, but have confirmed that our predictions do not vary markedly with other choices}. Importantly, we also allow this error term to act in an uncorrelated way on the separate bins of the analysis; this allows us to use the quadratic-in-dimension-6 term to get the characteristic size of the effects without assuming that the correlations predicted by that calculation must remain true after the inclusion of dimension-8 effects.

This treatment gives us a bin-by-bin variance estimate due to theoretical uncertainties of

\begin{equation}
    \sigma^2_{\rm th}=\sum_{\rm ops}\left(c_6^2N_{d6^2}+g_2^2c_8\sqrt{N_8}N_{d8}\right)^2,
\end{equation}
where the sum runs over all relevant exemplar operators, $c_6$ is the dimension-6 Wilson coefficient in question, $g_2$ is the $SU(2)$ gauge coupling, $N_{d6^2}$ is the number of events in the given bin predicted due to the quadratic dimension-6 operator effects, and
\begin{eqnarray}
    c_8=\sqrt{1+c_6^2},\\
    N_{d8}=\frac{1}{2}\sum_{F,B}N_{d6^2},
\end{eqnarray}
are our assumed value for an average dimension-8 Wilson coefficient and the distribution we use to simulate dimension-8 effects, carefully avoiding the assumption that their forward-backward behavior is well-described by that of the dimension-6-squared distribution. Note that this Wilson coefficient is perhaps a slightly aggressive choice, as the characteristic size of $c_8$ in the presence of a dimension-6 Wilson coefficient of $c_6$ should, by dimensional analysis and RGE considerations, actually be $c_8\sim c_6^2$; we adopt this slightly more aggressive approach as a compromise value to offset the added freedom we give to the $\mathcal{O}\left(\frac{1}{\Lambda^4}\right)$ distribution in treating it as uncorrelated between bins.

We assume that all of these errors are well-approximated by Gaussian distributions, and further that all these sources of error are completely uncorrelated. Our total error for a given bin then has the form

\begin{equation}
    \sigma^2_{\rm tot}=\sigma^2_{\rm stat}+\sigma^2_{\rm sys}+\sigma^2_{\rm th}.
\end{equation}

\subsection{Error Treatment in This Analysis: Specific Considerations}
\label{sec:asym}

The above generic error treatments allow us to simply define a test statistic as

\begin{equation}
    \chi^2=\sum_{\rm bins}\frac{\left(N_{\rm exp}-N_h\right)^2}{\sigma^2_{\rm tot}},
\end{equation}
where $N_h$ is the predicted number of events in a given bin for the hypothesis being tested.

In the absence of experimental data for the future detection reach projections derived here, we assume that the experimental data is well-described by the SM background, and thus our distribution can be simplified to give

\begin{equation}
    \chi^2=\sum_{\rm bins}\frac{N_{\rm sig}^2}{\sigma^2_{\rm tot}},
\end{equation}
where $N_{\rm sig}$ is the number of signal events predicted at $\mathcal{O}\left(\frac{1}{\Lambda^2}\right)$ for a given point in parameter space.

However, this treatment fails to exploit a useful feature of the distributions we're searching for in this case; the distinction between forward and backward contributions can be recast as an asymmetry, which will be less affected by systematic errors than any individual bin content.

Rather than simply binning in $\eta_{\ell^-}$ and performing a brute-force fit, as described above, we will fit instead two distributions simultaneously; the total cross section and the forward-backward asymmetry 

\begin{equation}\label{eq:asym}
    A_{\rm fb}\equiv\frac{N_F-N_B}{N_F+N_B},
\end{equation}
where $N_{F,B}$ is the number of events in a forward (backward) classification we will define in detail in \cref{sec:search},
each as a function of $M_{\ell\ell}$. While the information content of these two approaches, in terms of the number of terms in a $\chi^2$ sum, is identical, there is one major advantage to the asymmetry construction: due to the highly symmetric nature of our detectors at the LHC, the systematic errors on expected number of events are very strongly correlated under a swapping of the momenta of $\ell^+$ and $\ell^-$.

Because of this correlation, these systematic errors amount to a single uncertainty in the overall scale of both the numerator and denominator of \cref{eq:asym}, and therefore cancels out in the ratio\footnote{Of course, this cancellation would not be exact; in our analysis we only need assume it is accurate to the per-mille level to not materially impact our conclusions.}. The errors in this asymmetry then can be written as
\begin{equation}
    \sigma^2_{A_{\rm FB}}=4\frac{\sigma_{F,{\rm stat}}^2+\sigma_{F,{\rm th}}^2+\sigma_{B,{\rm stat}}^2+\sigma_{B,{\rm th}}^2}{\left(N_{F,{\rm bg}}+N_{B,{\rm bg}}\right)^2},
\end{equation}
where the systematic errors explicitly play no role in the case of the asymmetry variable. Our test statistic then becomes

\begin{equation}
    \chi^2=\sum_{M_{\ell\ell}~{\rm bins}}\frac{\left(N_{F,\rm sig}+N_{B,{\rm sig}}\right)^2}{\sigma^2_{F,\rm tot}+\sigma^2_{B,{\rm tot}}}+\frac{A_{\rm FB, sig}^2}{\sigma^2_{A_{\rm FB}}},
\end{equation}
where

\begin{equation}
    A_{\rm FB, sig}\equiv\left(1-A_{\rm FB, bg}\right)N_{F,{\rm sig}}-\left(1+A_{\rm FB, bg}\right)N_{B,{\rm sig}}
\end{equation}
is the signal contribution to the forward-backward asymmetry, linearized in $\frac{1}{\Lambda^2}$ as required for consistent treatment of the EFT perturbation series.

\section{Searching in Dileptons at the LHC}
\label{sec:search}

Our analysis begins with parton-level Monte Carlo events generated by \texttt{MadGraph5}~\cite{Alwall:2014hca} with the SMEFTsim package~\cite{Brivio:2017btx,Aebischer:2017ugx}, with both the renormalization and factorization scales chosen dynamically to be the sum of the transverse energies of the produced leptons. We also utilize \texttt{Pythia8}~\cite{Pythia1} and \texttt{Delphes}~\cite{deFavereau:2013fsa} to shower, hadronize, and simulate the detector response to our events. Our analysis was constructed and applied using the \texttt{AEACuS} package~\cite{walker}\footnote{While this package was originally constructed to run on LHCO-format events, it has since been expanded to construct LHCO-like files which are able to treat event-by-event weights consistently, which was essential to this analysis.}.

Our analysis is inspired by an ATLAS result \cite{Aaboud:2017buh} searching for new physics effects in dilepton production in $36.1\,{\rm fb}^{-1}$ of data; we choose to follow their binning in dilepton invariant mass $M_{\ell\ell}$ to utilize their reported systematic errors. Our bin edges in $M_{\ell\ell}$ are thus:
\begin{align}
    \left(80,120,250,400,500,700,900,1200,1800,3000,6000\right).
\end{align}
A finer binning, or indeed an unbinned analysis, may be more natural at greater integrated luminosities such as those we explore here, but we do not have access to reliable systematic error treatments in that context, so we do not consider them further. For our initial event selection cuts, we require exactly one same-flavor, opposite-sign lepton ($e$ or $\mu$) pair, each of which has $|\eta| < 2.47$ (but avoiding the transition region $1.37\leq|\eta|\leq1.52$ to maintain designed energy resolution) and a transverse momentum of $P_T \geq 10$~GeV. We veto events which contain a jet with $|\eta| \leq 2.5$ and $P_T \geq 30$~GeV.

Following event selection, we bin in $M_{\ell\ell}$ as described above, and also bin events into forward and backward categories. The forward bins are defined to contain all events where the (negatively charged) lepton is more forward than the anti-lepton, i.e. $|\eta_{\ell^-}|>|\eta_{\ell^+}|$. The operator class labels of course do not guarantee that every event falls into the respective bin (as seen explicitly in \cref{fig:templates}), but they do preferentially populate the associated bin, e.g. samples generated using forward exemplars (whether direct or shift) preferentially, but not exclusively, populate the forward bin as defined here. The strength of this preference is affected by the parton distribution functions and grows with increasing $M_{\ell\ell}$, as the true but unmeasurable partonic forward direction is more likely to coincide with this definition when one incoming parton has a larger fraction of the proton's momentum, increasing the likelihood that it is a valence quark.

The SM background normalization is adopted from the analysis in \cite{Aaboud:2017buh} to account for the higher-order and detector effects included in that analysis; we retain the leading-order forward-backward behavior taken from the \texttt{MadGraph5}, \texttt{Pythia8}, and \texttt{Delphes} Monte Carlo chain out of necessity, as the angular behavior is not reported in \cite{Aaboud:2017buh}; corrections to this value due to higher order SM effects can be included without harming the sensitivity of our analysis. The normalization factor needed to agree with experimental background rates is retained as a ``k-factor'' to account for imperfections of our Monte Carlo chain, most notably with respect to detector simulation; it remains of order 1 for dimuon events in every bin of the analysis, with a greatest value of 2.5; the ATLAS result~\cite{Aaboud:2017buh} reports more dielectron events than dimuon events, in contrast to the Delphes default simulation, so the ``k-factor'' for those gets a bit higher, up to 6.7.

To simulate our signal and error samples we utilize exemplar operators as discussed in \cref{sec:shifts}, simulating each distribution one at a time and exploiting the linearity of our signal function in Wilson coefficients to construct a complete signal hypothesis as a scaled sum of the four samples we simulated. All signal and error distributions are scaled by the efficiency-correcting ``k-factor'' found from the background construction, as the detector effects are insensitive to the internal structure of the hard scattering process. Error distributions are treated similarly to signal distributions; we assume that interferences between two distinct exemplar distributions are not important to the description of the characteristic error size and combine the error distributions linearly after applying the scaling described in \cref{sec:stats} to each distribution individually. This non-interference assumption is perfectly physical when applied to the interference between forward and backward distributions due to the distinct helicities involved in the initial and final states of those scatterings; it is a reasonable approximation in the case of interfering shift- and direct-type operators due to their preferences for opposite ends of the $M_{\ell\ell}$ distribution.

\begin{figure}[t]
    \center
    \subfloat[\label{direct}Direct Forward and Backward]{{\includegraphics[width=0.4\textwidth]{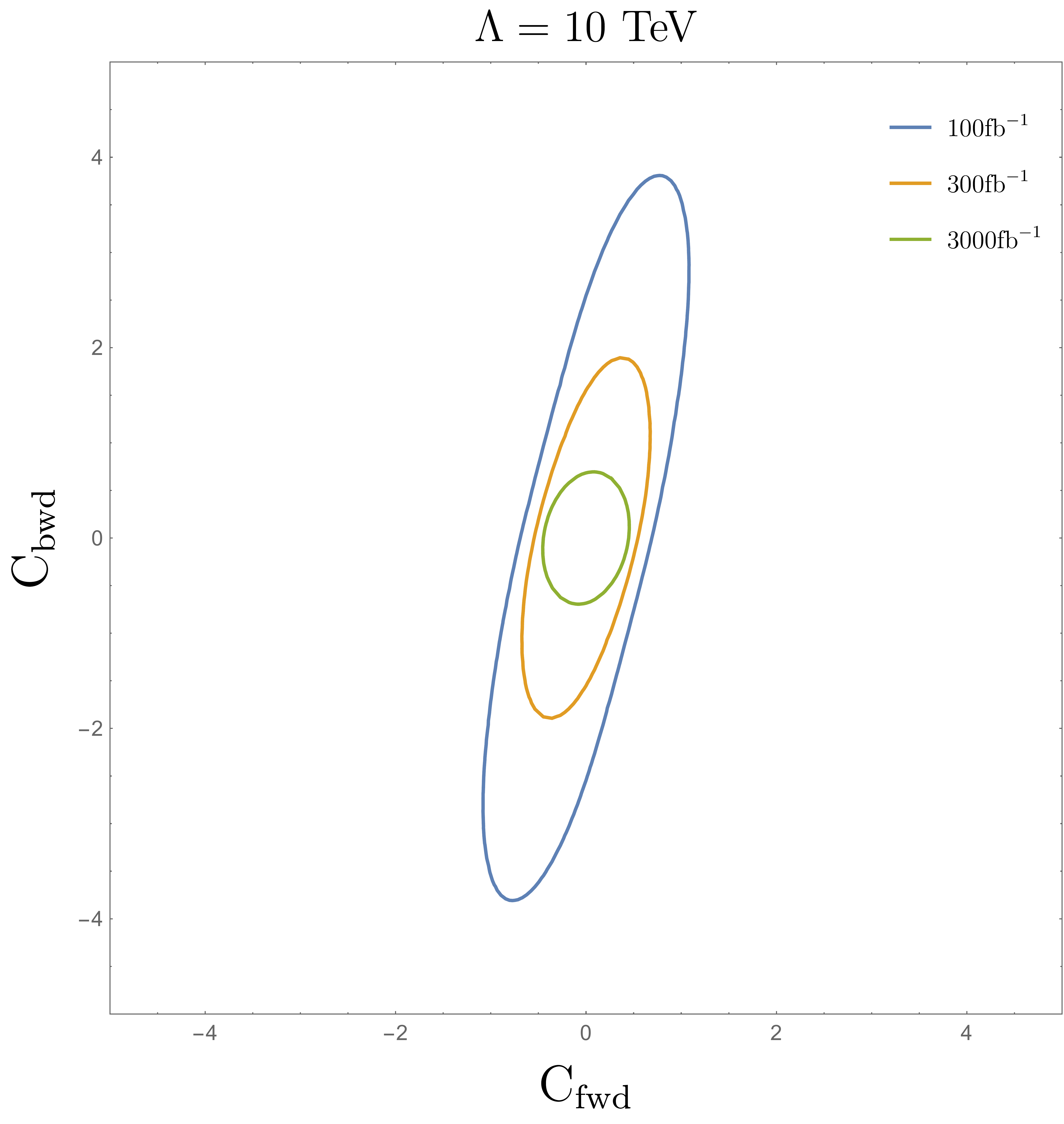}}}\hspace{1em}
    \subfloat[\label{forward}Direct and $Z$-Pole Shift Forward]{{\includegraphics[width=0.4\textwidth]{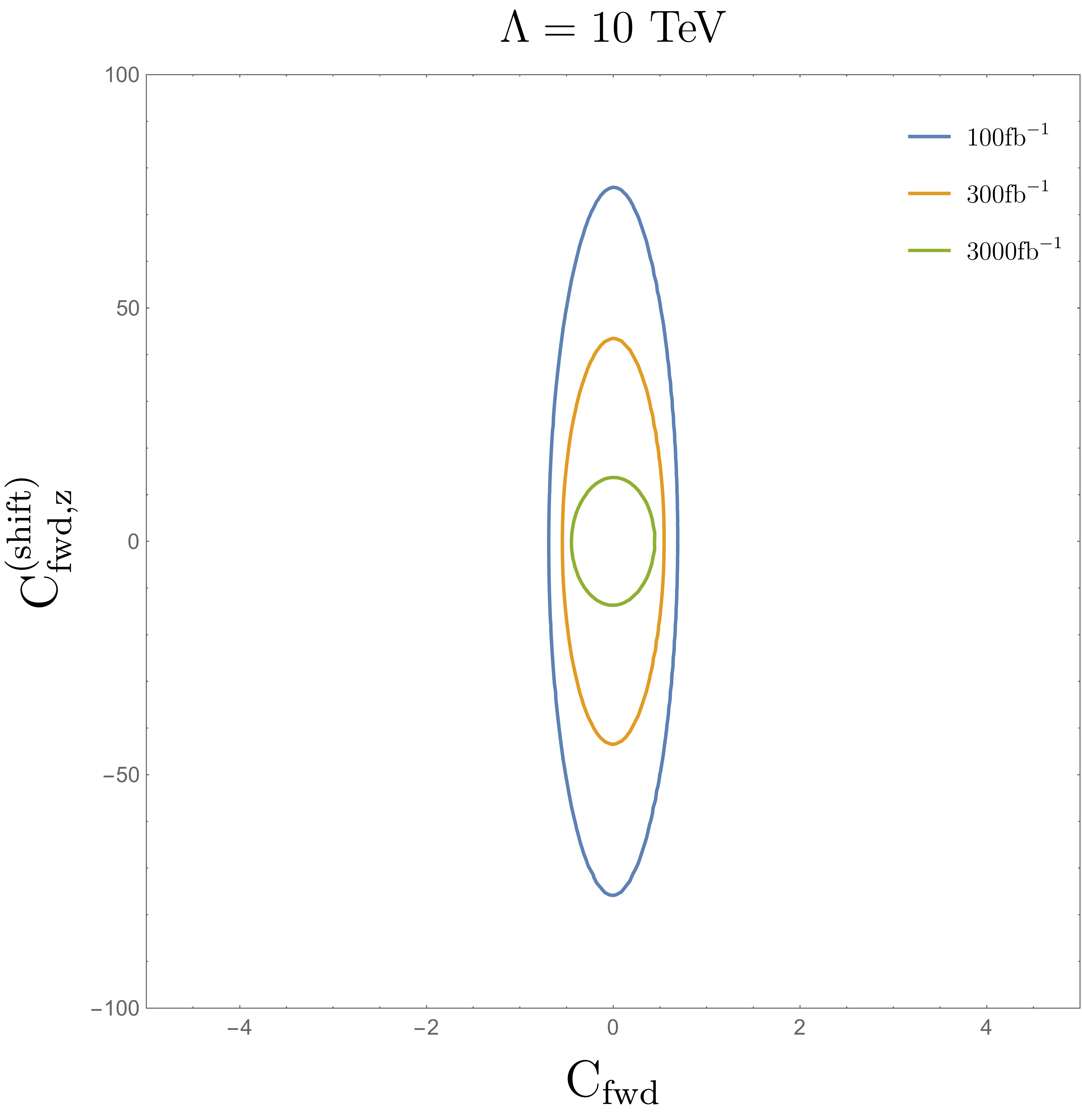}}}
    \hfill 
   
   \center
    \subfloat[\label{backward}Direct and $Z$-Pole Shift Backward]{{\includegraphics[width=0.4\textwidth]{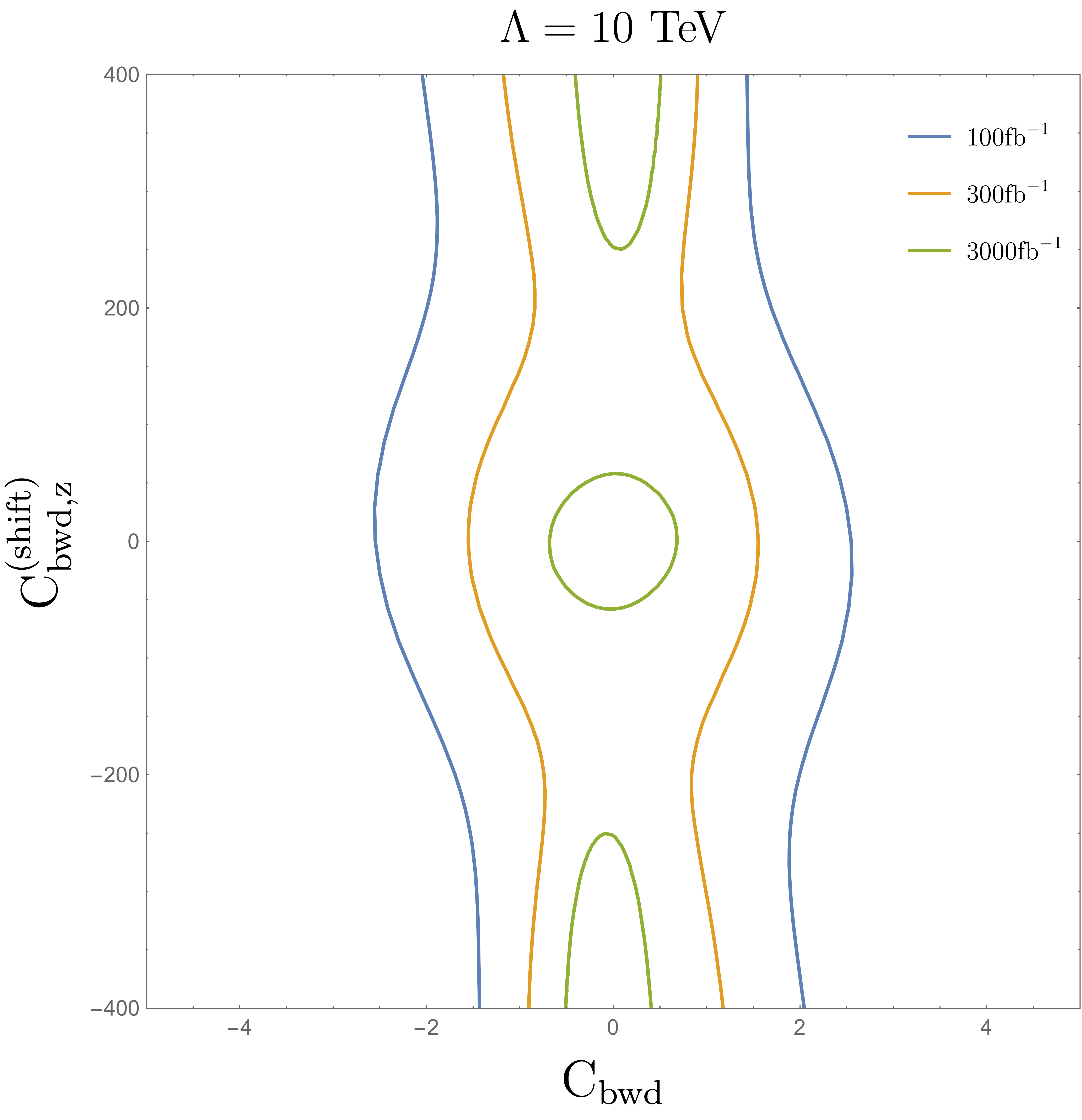}}}\hspace{1em}
    \subfloat[\label{shift}$Z$-Pole Shift Forward and Backward]{{\includegraphics[width=0.4\textwidth]{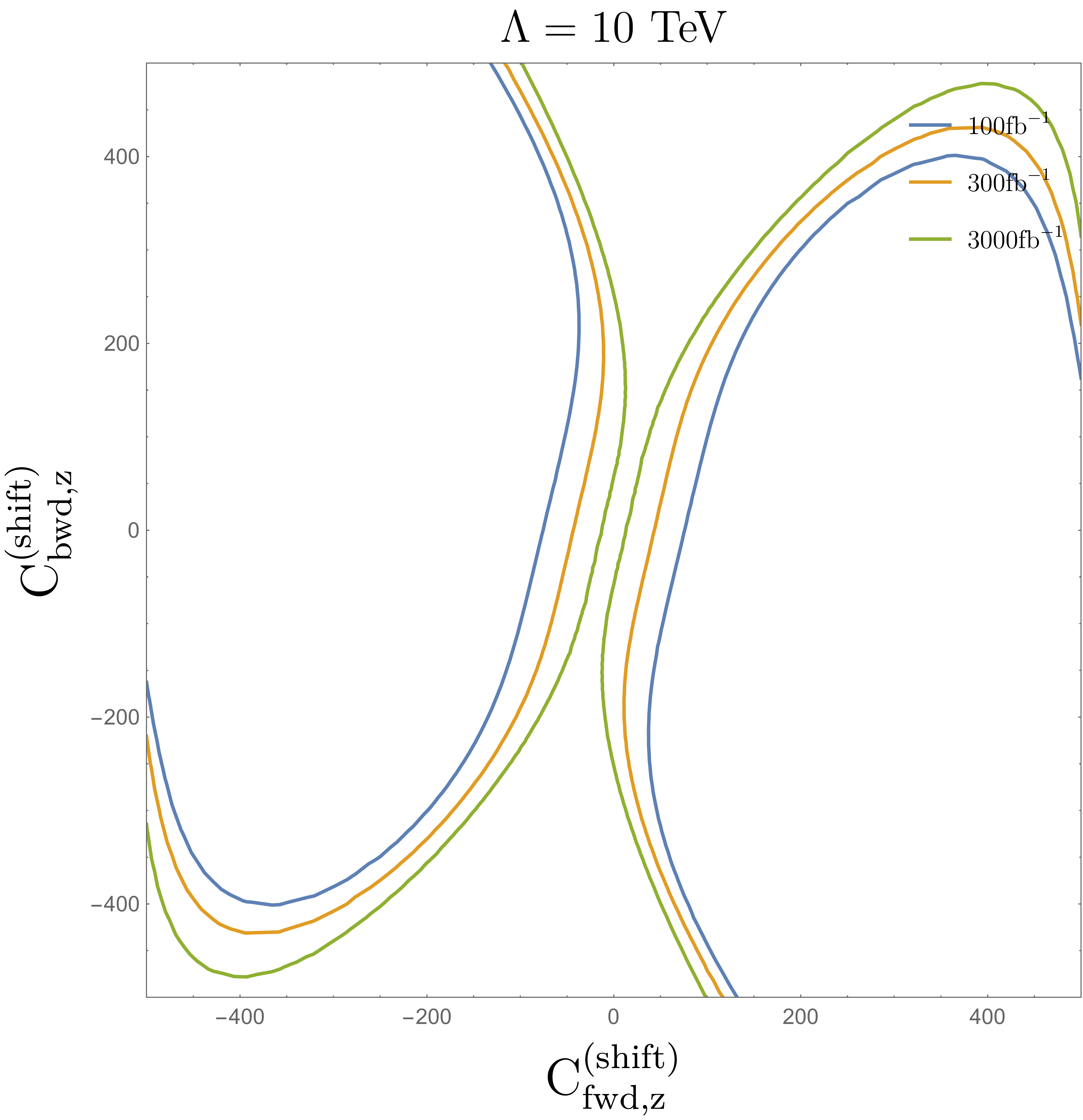}}}
    \hfill
    \caption{\label{fig:bounds}Constraints on SMEFT parameter space arising from this analysis. All plots show slices along the plane of vanishing Wilson coefficients not shown. The blue, orange, and green curves correspond to the projected constraints based on performing this analysis using 100, 300, and 3000 fb$^{-1}$ of data.}
\end{figure}

Our high-energy shift signal functions require one additional element of care, as we find it impossible to perfectly tune away contributions to the $Z$-pole bin through our template fitting techniques. We therefore manually discard contributions to the lowest $M_{\ell\ell}$ bin from those two signal distributions. Our error estimates are not subjected to this procedure, as even a perfect tuning at $\mathcal{O}\left(\frac{1}{\Lambda^2}\right)$ does not imply anything about cancellations at higher orders in the SMEFT perturbation series.

Performing the statistical analysis as described in \cref{sec:stats}, we find the regions of Wilson coefficient space which are constrained at 95\% C.L. for various amounts of integrated luminosity and show those curves in \cref{fig:bounds}. We are unable to constrain either of the high-energy shift directions using this data, so we display plots only for the remaining four directions in Wilson coefficient space.

The direction of the most stringent constraint depends on the nature of the operators under consideration. For direct, energy-growing contributions, the direction of most rapid change in the cross section is generically best constrained, because there are so few events in the high-energy bins that the uncertainty on the forward-backward asymmetry in the data is large. By contrast, for the Z-pole shift effects, the asymmetry is the sole source of constraint because the background is very large; this allows a precise asymmetry measurement (where we anticipate cancellation of systematic errors in our error treatment, as discussed in \cref{sec:asym}), but gives a very large statistical and systematic error on the total cross section. This explains why only one particular combination of forward and backward contributions is constrained by our data.

In \cref{direct}, we were able to get a closed and tight bound on the direct four-fermion operator contributions; these bounds are actually slightly stronger than those in \cite{Alte:2018xgc}, due to the inclusion of both electron and muon final states in the analysis. This indicates that the additional errors arising due to shift operators do not have significant impact on constraints for energy-growing phenomena in this final state.\footnote{The generalizability of this result to other final states is far from obvious.} The closure of the ellipse in the elongated direction is due to the asymmetry construction; without the cancellation of systematic errors this direction is very difficult to constrain, as it corresponds to an unchanging total cross section and increasing theoretical errors. 

In \cref{forward} we note that the constraint ellipse is oriented perfectly along the axes of the plot, indicating that the two operators do not materially affect one another. This is due to their primary effects being in bins on opposite ends of the $M_{\ell\ell}$ distribution. The most clear impact of the systematic error cancellation here is the significant shrinking of the ellipse in the vertical direction with added data; the outermost ellipse already corresponds to constraints in the lowest $M_{\ell\ell}$ bin which, if constructed binwise rather than using $A_{FB}$, would be of order the systematic error floor of 2\%, and thus would be impervious to improvement with increased statistics.

\cref{backward} is similarly aligned along the axes of the plot, but note that there is slightly more interplay in the case of backward contributions than forward ones, due primarily to the slightly smaller magnitude of the cross section in the direct backward case relative to the forward case. Note also the differing normalization of the vertical axis with respect to \cref{forward}; with HL-LHC luminosities the constraints eventually close in the vertical direction, albeit still at quite large values. The impact of theoretical errors on this plot is also clear in the re-emergence and/or widening of allowed region at notably larger Wilson coefficients.

The constrained regions shown in \cref{shift} are due entirely to the forward-backward asymmetry; significant impacts of changes in the total cross section would occur only at Wilson coefficients of order thousands, clearly outside the range of applicability of the EFT approach. In this plane it is clear that there is a direction which is unconstrained, as there must be. The diagonal unconstrained direction corresponds to EFT effects which change only the total cross section and not the forward-backward behavior of dilepton events at the Z-pole. Once again, note that the allowed region widens significantly at large Wilson coefficients due to the impact of theoretical uncertainties. Our bounds on $Z$-pole shift contributions from the SMEFT are competitive with bounds arising from low-energy precision data \cite{Berthier:2016tkq}. Projecting the precision data constraints into this plane is difficult, and the interplay between theoretical errors at high and low energies is not yet fully understood, but we have verified that some points which are allowed under the most restrictive methods applied (with large caveats about their overly aggressive nature) in \cite{Berthier:2016tkq} are constrained by these results already with just 100 fb$^{-1}$ of data.

\section{Results and Outlook}
\label{sec:results}

We have demonstrated that it is possible to derive bounds on both direct, four-fermion operator contributions and shift-type operator contributions to dilepton production at the LHC simultaneously. The simultaneous analysis does not significantly dilute the strength of bounds on energy-growing four-fermion interactions as explored in \cite{Alte:2018xgc} (and improved on here by the inclusion of additional data in the form of muonic events), and the bound on shift type operators from the LHC forward-backward asymmetry are competitive with those arising from LEP and other lower-energy data. We note also that the particular linear combinations of operators which contribute to forward and backward shift-type contributions at the LHC are distinct from those at LEP due to the parton distribution function effects of the proton-proton initial state, making these constraints complementary to those which can be derived from $Z$-pole studies.

This furthers the ultimate goal of developing sufficient observables, calculated consistently in the SMEFT, to constrain a reasonably-defined parameter space of the theory, by giving two new directions of constraint, corresponding to the SM parameter shift directions, which are linearly independent from those bounded by previous measurements, and updating the constraints on direct four-fermion operator contributions to dilepton production at the LHC. The eventual output of work in this vein must be a global fit of all available data to the SMEFT which yields a region in Wilson-coefficient space that is ruled out by the totality of measurements made, which will be of use in directing future model building and experimental efforts towards the regions which are less constrained by the set of precision measurements already made.

Producing those bounds in a way that is conservative enough to make them, at worst, much more difficult to evade by creative model-building will lead to a tool of utility comparable to the output of the LEP Electroweak Working Group.  Such a construction is able to serve a first line of analysis for new models in comparing their effect on precisely-measured SM processes to the data at multiple energy scales with minimal work on the part of the model builder. Utilizing theory errors of the type employed here is central to this becoming a broadly-used tool rather than a result which is of interest only to its immediate practitioners.

\section*{Acknowledgments}
The authors are grateful for multiple discussions with Matthias K\"onig during the completion of this article. A.~Horne, J.~Pittman, M.~Snedeker, and J.~W.~Walker acknowledge support from NSF grant PHY-1820801.

\section*{Note Added:}

As this article was nearing completion, Ref. \cite{Aad:2020otl} appeared. This search utilizes significantly more data, but is essentially predicated on treating the $\mathcal O \left(\frac{1}{\Lambda^4}\right)$ terms, discussed here as a probe of theoretical uncertainties in the SMEFT, as part of the signal function. This choice renders the study much less usable than the previous \cite{Aaboud:2017buh} for producing theoretically consistent bounds on the SMEFT.

\bibliography{Sources}

\end{document}